\documentclass[%
 reprint,
floatfix,
superscriptaddress,
showpacs,preprintnumbers,
 amsmath,amssymb,
 aps,
prl,
]{revtex4-1}
\usepackage{xcolor}
\usepackage{times}
\usepackage{amssymb}
\usepackage{dsfont}
\usepackage{graphicx}
\usepackage{dcolumn}
\usepackage{bm}
\usepackage{epstopdf}
\usepackage{lineno}
\usepackage[colorlinks]{hyperref}
\hypersetup{colorlinks = true,
linkcolor = blue,
filecolor = blue,
urlcolor = blue,
citecolor = blue}
\begin{document}

\title{Charge Modulation in the Vortex Halo of a Superconductor Enhances its Critical Magnetic Field}

\author{Anurag Banerjee}
\affiliation{Indian Institute of Science Education and Research Kolkata, Mohanpur, India-741246}
\affiliation{Universit\'{e} Paris-Saclay, Institut de Physique Th\'eorique,  CEA, CNRS, F-91191 Gif-sur-Yvette, France}

\author{Catherine P\'epin}
\affiliation{Universit\'{e} Paris-Saclay, Institut de Physique Th\'eorique,  CEA, CNRS, F-91191 Gif-sur-Yvette, France}

\author{Nandini Trivedi}
\affiliation{Department of Physics, The Ohio State University, Columbus, Ohio 43210, USA}

\author{Amit Ghosal}
\affiliation{Indian Institute of Science Education and Research Kolkata, Mohanpur, India-741246}

\begin{abstract}
When an orbital magnetic field suppresses superconductivity, forming periodic vortices in type-II superconductors, subdominant orders can emerge in the vortex cores. Rather than competing with superconductivity, we find that the emergent charge order within the halo of a vortex makes superconductivity more robust by enhancing the upper critical field. We establish that charge modulations nucleate in and around the vortex core for model parameters dictated by the underlying non-superconducting state. We further show that the spectral signatures from the Caroli-de Gennes-Matricon (CdGM) bound states in vortex cores track the charge modulation. 
The CdGM-like peak is found to shift toward the gap edge and oscillate from particle-to-hole bias from site to site, signaling charge modulation.
\end{abstract}

\maketitle

\noindent \textit{Introduction} -- 
Although there has been much discussion of competing orders such as magnetic order, charge order, and pair-density wave order~\cite{Dagotto, Norman196, Cava0, Lake1759, Chang2012, Wise2008, silvaNeto2014, PressurePRL} for strongly correlated superconductors (SC), here we investigate the surprisingly less studied classic problem: the evolution of vortices in an s-wave superconductor proximate to a charge-ordered state.  
Our main finding is that in an externally applied magnetic field, the vortex core shows strong charge modulations, instead of the expected metallic Abrikosov vortex~\cite{AVL,HessSTM}, which provides additional stability to the mixed state of SC against the increasing magnetic field, and {\it enhances} the upper critical field, depicted in Fig.1. A distinct signature of the charge-modulated vortex core is shown to arise from the energy and location of in-gap states within the vortex core, constituting the so-called Caroli-de Gennes-Matricon (CdGM) peak.

\begin{figure}[t]
\includegraphics[width=0.49\textwidth]{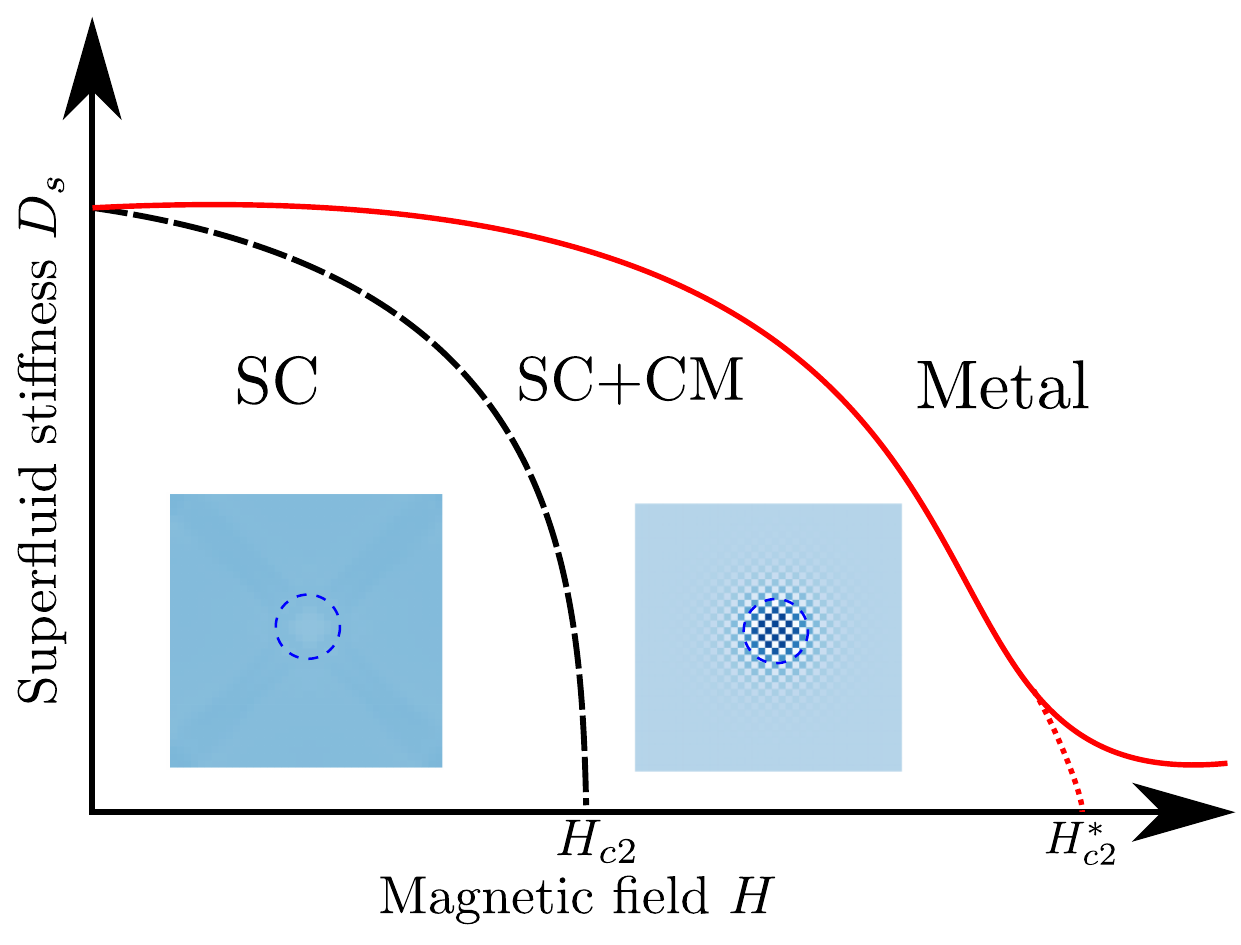}
\caption{The charge-modulated vortex cores provide additional stability to the mixed state of SC as $H$ increases, compared to a conventional vortex state with metallic cores. When the core (region blue circle in the inset) has a uniform electronic density, the superfluid stiffness, $D_s$, vanishes at $H_{c2}$. Instead, when charge modulations germinate around the vortex, $D_s$ survives until $H^*_{c2} > H_{c2}$. 
}
\label{Schematics}
\end{figure}

Vortex halos that reveal subdominant orders locally around the core~\cite{GuillamonSTM, Hoffman466, Kalinsky, Anushree_dVortex, SachdevPRB, TKLeeVortex, Dai18, AFM_Arovas, DebmalyaPRB, Wang2001, Lee2000, Zhang2002,mahato2024, GaneshPRB, GaneshJPSC,Wen2023,Du2020,PDWNbSe2, EdkinsVortex,Zhang1089, PRBtJ_22,CDW_COM, Rini_PRB, AgterbergPDW, LeePRX14} have been reported previously but the enhancement of $H_{c2}$ that we find in ordinary SC proximate to charge-ordered states is new (shown schematically in Fig.~\ref{Schematics}). 
We expect the mechanism that leads to the enhancement of $H_{c2}$ in conventional SC to be relevant to other superconductors as well, and hope that our results will guide experiments on cuprates~\cite{Hoffman466,EdkinsVortex,AnomalousVLCuprates,Wu2013,Hc2Cuprates,MingCdGM,Dragana,DisorderCDWHTSc,Thomarat2024,Zou2022,Ye2023} and flat-band moire materials~\cite{Cao2018,Xia2025,Guo2025,Tian2023,TuningTBG,Cao2018_2,Lu2019}.

We focus here on the attractive Hubbard model (AHM) that supports degenerate charge density wave (CDW) and s-wave superconductivity strictly at half-filling~\cite{PhaseAHM, MicnasRMP}. Perturbing the system slightly away from half-filling destabilizes the CDW order while keeping superconductivity robust~\cite{Moreo}. Past studies have shown the evolution of CDW order by including vortices at half-filling~\cite{GaneshPRB, GaneshJPSC}. Here, we study the system away from half-filling, where SC dominates, and demonstrate that charge modulations (CM) can still emerge in the vortex cores in the mixed state of such an SC. We establish the conditions for the nucleation of CM around the core regions and explore how conventional signatures in spectroscopic probes~\cite{HessSTM, CdGM,JCDavisLDOS1} alter due to subdominant CDW and closely associated pairing modulations. Henceforth, we refer to such vortex cores as `CM-cores', as depicted in Fig.~\ref{Schematics}. Furthermore, we show that such CM-cores can protect the superconductivity against the magnetic field, $H$.

\bigskip

\noindent \textit{Model and methods} --
We consider the attractive Hubbard Model (AHM) in the presence of an orbital magnetic field: 
\begin{equation}
\mathcal{H}=-t\sum_{\langle i,j\rangle,\sigma} e^{i\phi_{ij}}c_{i\sigma}^{\dagger} c_{j\sigma}-\vert U \vert \sum_i \hat{n}_{i\uparrow}\hat{n}_{i\downarrow} -\mu \sum_{i,\sigma} \hat{n}_{i\sigma}\
\label{ahm}
\end{equation}
where $c_{i\sigma}(c_{i\sigma}^\dagger)$ annihilates (creates) an electron with spin $\sigma$ at site $i$ of a two-dimensional square 
lattice, and $n_{i \sigma}=c^{\dagger}_{i\sigma} c_{i\sigma}$ is the local density operator with spin $\sigma$. The first term represents electrons hopping between nearest neighbors $\langle ij\rangle$, where the Peierls phase factor $\phi_{ij}=\frac{\pi}{2\Phi_0}\int_i^j \mathbf{A}.\mathbf{dl}$ represents the orbital magnetic field -- here, $\Phi_0=hc/2e$ is the superconducting flux quantum. We choose Landau gauge: $\mathbf{A}=Hx\hat{y}$, which produces a constant uniform magnetic field $H$ in the $\hat{z}$ direction. $H$ is expressed in the units of $H_0$, the magnetic field strength for flux $\Phi_0$ passing through a system size of $40 \times 40$. We tune the magnetic field by changing the size of the system that accommodates the flux $\Phi_0$.

The second term in Eq.~({\ref{ahm}}) is the on-site attraction of strength $\vert U\vert$ between the electrons leading to sSC and CDW orders at the level of mean field theory. We perform Bogoliubov-de Gennes (BdG) calculations~\cite{PdGbook,GTR,Datta_swave} amounting to self-consistent determination of local charge density $\rho_i=\sum_{\sigma}\langle c^{\dagger}_{i\sigma} c_{i\sigma} \rangle$, and sSC pairing amplitude $\Delta_i=-U \langle c_{i\downarrow} c_{i\uparrow} \rangle$ at all sites. 
We consider
$\rho_i=\rho+\Delta_{\rm CDW}(i) \cos(\mathbf{Q}.\mathbf{r})$~\cite{AnuragPRB,Proximity} with $\mathbf{Q}=(\pi,\pi)$.
The chemical potential $\mu$ fixes the average density of electrons $\rho=N^{-1}\sum_i n_{i \sigma}$. We use a magnetic unit cell (MUC) containing two superconducting flux quanta and extend the wave function of the whole system containing up to $200$ MUCs using a repeated zone scheme~\cite{Kalinsky} so that our vortex lattice contains up to $400$ SC-vortices. All our calculations are set at zero temperature ($T=0$), and energies are measured in units of $t=1$, and lattice spacing is considered unity. More details are included in the supplementary materials (SM)~\cite{Supple}.
\vspace{1.5mm}

\begin{figure}[t]
\includegraphics[width=0.45\textwidth]{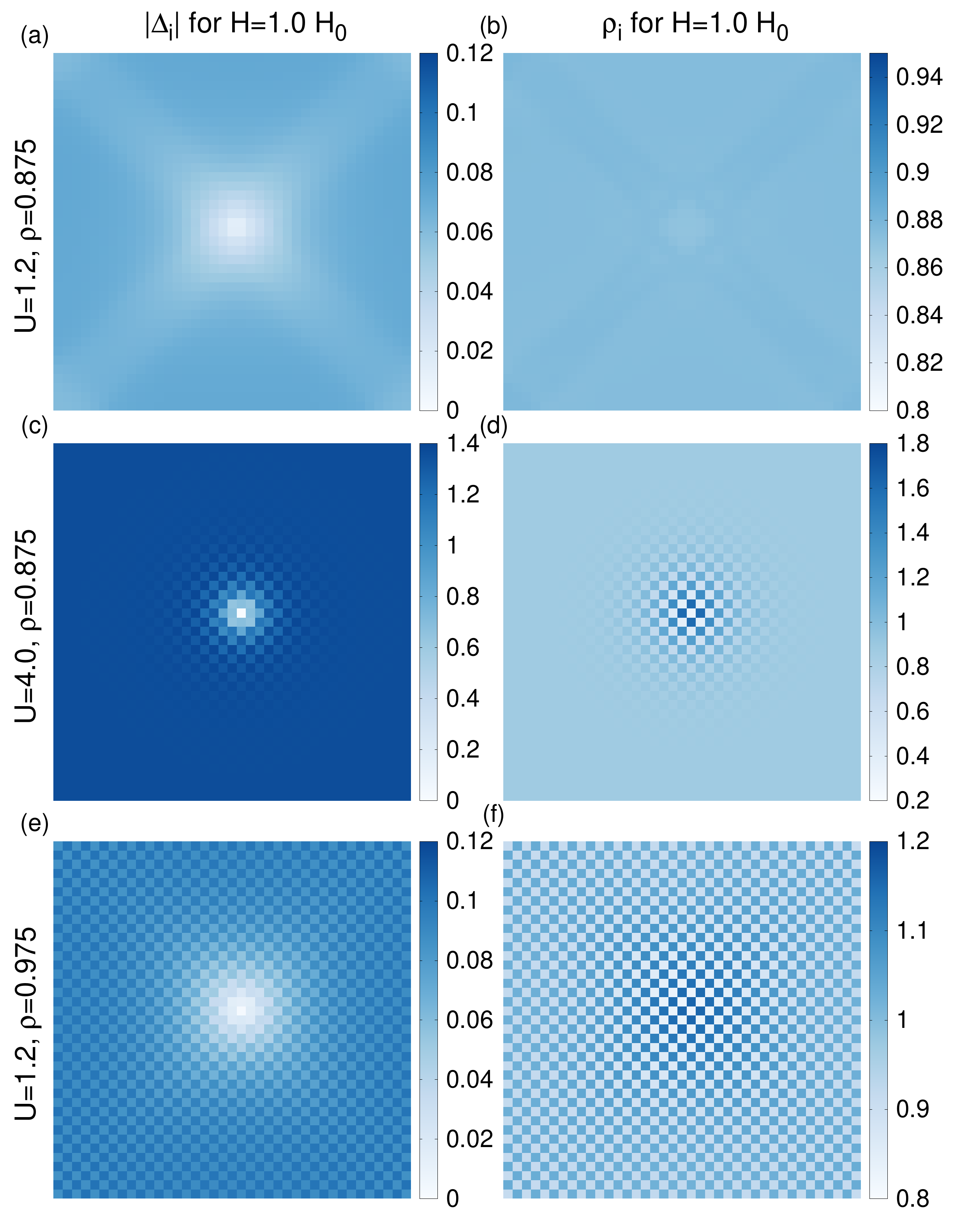}
\caption{
The self-consistent spatial profile of $\Delta_i$ (right panel) and $\rho_i$ (left panels) is shown for three sets of ($U$,$\rho$) in top, middle, and bottom panels for $H=H_0$. \textbf{(a-b)} results in spatially uniform density in the vortex-core region of depleted SC for $(1.2,0.875)$. \textbf{(c-d)} CM-cores are generated for $(4.0,0.875)$ -- the CM depletes rapidly beyond $\xi_{\rm SC}$. $\Delta_i$ also shows modulations within cores. $\textbf{(e-f)}$ The length-scale of modulation enhances beyond $\xi_{\rm SC}$, marking a `vortex-halo' in $\Delta_i$ and $\rho_i$ for $(1.2,0.975)$.    
}
\label{spatial}
\end{figure}

\noindent \textit{Spatially modulated order parameters:}
We present in Fig.~\ref{spatial} the spatial profile of the pairing amplitude, $\Delta_i$, and density, $\rho_i$, for different parameters ($U,\rho$) in Eq.~({\ref{ahm}}). Here, we distinguish the behavior in two regimes by tuning the parameters of the AHM. The weak coupling regime $U\lesssim t$ in which the pairing energy gap scale is much smaller than the superfluid stiffness, and the strong coupling regime $U\gtrsim t$ with a reversal of these two energy scales~\cite{GTR,GTR_PRB}. 

For a single vortex, Fig.~\ref{spatial}(a, b) shows the top-view of a standard conical-shaped vortex and a near-uniform density profile for $(U,\rho)=(1.2,0.875)$. Next we see that spatial modulation in local orders with $\bf{Q}=(\pi,\pi)$ nucleates around vortex-core under two circumstances: by increasing the coupling keeping $\rho$ unaltered, e.g., for $(U,\rho)=(4,0.875)$ in Fig.~\ref{spatial}(c, d), or, by tuning the density toward half filling ($\rho=1$) even for a weakly coupled superconductor $(U,\rho)=(4,0.975)$, as shown in Fig.~\ref{spatial}(e,f). Notice that the length scale of CM strongly depends on $(U,\rho)$. The SC-coherence length $\xi_{\rm SC} \sim \Delta^{-1}$ is the only length-scale in pristine SC (Fig.~\ref{spatial}(a, b)) that governs the variations near a vortex core. The emerging CM brings another length scale $\xi_{\rm CM} \sim \Delta_{\rm CDW}^{-1}$. The presence of competing orders interrelates these two length scales and modulates the density and order parameter, as shown in Fig.~1(c-f); simple estimates of $\xi_{\Delta}$ and $\xi_{\rm CM}$ are given in the SM~\cite{Supple}. In particular, for ($U,\rho$) combinations that favor CM, we find $\xi_{\rm CM} > \xi_{\rm SC}$. Thus, our estimates naturally explain how charge modulations exist even outside a vortex core in a region characterized by $\xi_{\rm CM}$ (termed `vortex halo') consistent with   experiments~\cite{Hoffman466, Wu2013}.

Away from half-filling, a spatial modulation in $\rho_i$ accompanies the corresponding modulation in $\Delta_i$~\cite{AnuragPRB}, and recent experiments extracted signatures of a modulated pairing in the vortex halo~\cite{PDWNbSe2, Edkins976}. The emergence of CM in the vortex halo away from half-filling is intriguing, and it is not an effect of strong-coupling alone. We illustrate this in Fig.~\ref{spatial}(e,f) by presenting qualitatively similar results for $(U,\rho)=(1.2,0.975)$, a relatively weaker coupling but for a density closer to half-filling. Not only do the CMs around vortex cores persist, but the modulations become longer-ranged in this case.

Remarkably, the average electron density around the core, $\tilde{\rho}$, approaches {\it local} half-filling whenever CM nucleates, regardless of the average density of the entire system, $\rho$. In contrast, $\tilde{\rho}\approx \rho$ remains near-homogeneous in the absence of CM. More details are provided in SM~\cite{Supple}.

The natural question arises: Can we predict which ($U, \rho$) generates CM-cores? Crucial insight into this question is provided by the underlying `normal' state for these parameters, which we discuss below.

\begin{figure}[t]
\includegraphics[width=0.49\textwidth]{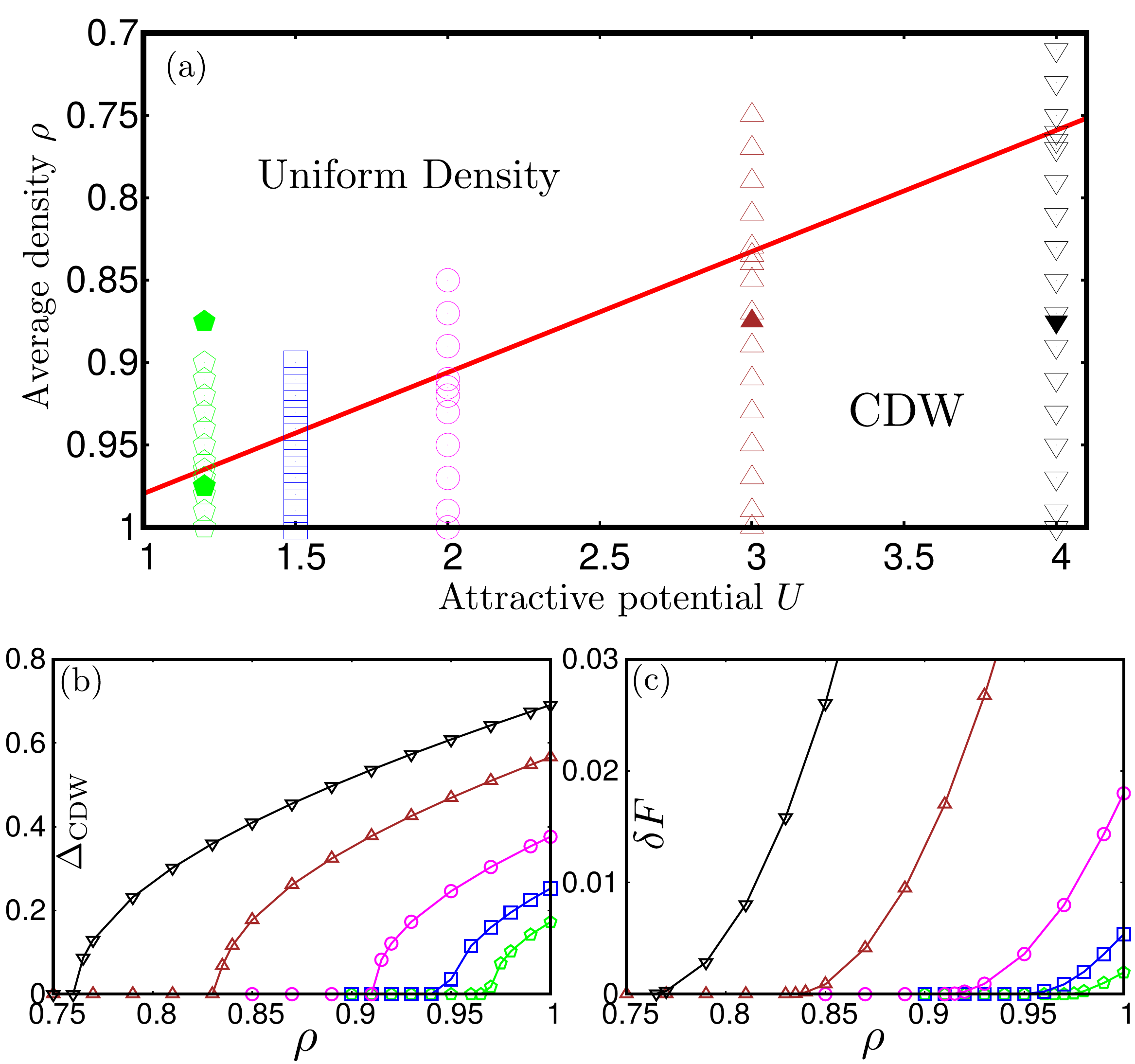}
\caption{\textbf{(a)} Phase diagram of the normal state of AHM of Eq.~({\ref{ahm}}) in the $U$–$\rho$ plane. A uniform CDW state is energetically favored over the uniform Fermi liquid up to a critical $\rho_c(U)$, and the red line
divides the phase space.
Empty points indicate parameters for normal state calculations (without pairing), while the filled markers also include the vortex calculation upon including pairing and $H$. The color of markers identifies specific values of $U$. The same color convention is used for panels (b) and (c).\textbf{(b)} Evolution of Average CDW amplitude $\Delta_{\rm CDW}$  with $\rho$. \textbf{(c)} Free energy difference $\delta F\equiv F_{\rm uni} - F_{\rm CDW}$  as a function of $\rho$. The red line in panel (a) is a linear fit of $\rho_c(U)$ at which $\delta F$ and $\Delta_{\rm CDW}$ vanish.
}
\label{PhaseDiag}
\end{figure}

\vspace{3.5 mm}
\noindent \textit{Charge modulation in the {\it normal} state:}
We now extract the phase diagram of the underlying {\it normal} state of AHM in the absence of $H$. The normal state is the mean-field solutions of ${\cal H}$ in Eq.~(\ref{ahm}) upon suppressing Cooper-pairing (i.e., $\Delta_i = 0$ for all sites $i$, but retaining the Hartree terms~\cite{GTR_PRB,ChakrabortyPNS}. Although uniform CDW and SC are energetically degenerate at $\rho=1$, superconductivity dominates away from this particle-hole symmetric point. However, a subdominant CDW order emerges near half-filling once the pairing is suppressed, as shown in Fig.~\ref{PhaseDiag}(a). An increase in $U$ enhances the phase space of CDW. Fig.~\ref{PhaseDiag}(b) illustrates the reduction of the amplitude of CDW, $\Delta_{\rm CDW}$, by tuning $\rho$ away from the half-filling for different $U$, indicating that $\Delta_{\rm CDW}$ vanishes at a $U$-dependent critical density $\rho_c(U)$. Similarly, the difference between the ground-state energies of the uniform metallic phase, $F_{\rm uni}$, and the CDW phase, $F_{\rm CDW}$, also vanishes at the same $\rho_c(U)$, as shown in Fig.~\ref{PhaseDiag}(c).

In the aforementioned normal state, we assumed that the amplitude of charge modulations $\Delta_{\rm CDW}$, whenever present, is spatially uniform. Interestingly, for $\rho \lesssim 1$ we have found signatures that the lowest energy charge-modulated state is the one that forms a large CM-puddle, with the puddle-averaged electronic density anchored at half-filling.
Outside of CM-puddles, typically $\rho_i \leq \rho$. Our picture of a phase-separated
CM-puddles partitioned by regions of near-homogeneous density persist over a considerable range of $(U,\rho)$, and lead to half-filled CM-cores seen in Fig.~\ref{spatial}(d, f).
Charge-modulated and inhomogeneous normal states are discussed further in the SM~\cite{Supple}.

\begin{figure}[t]
\includegraphics[width=0.40\textwidth]{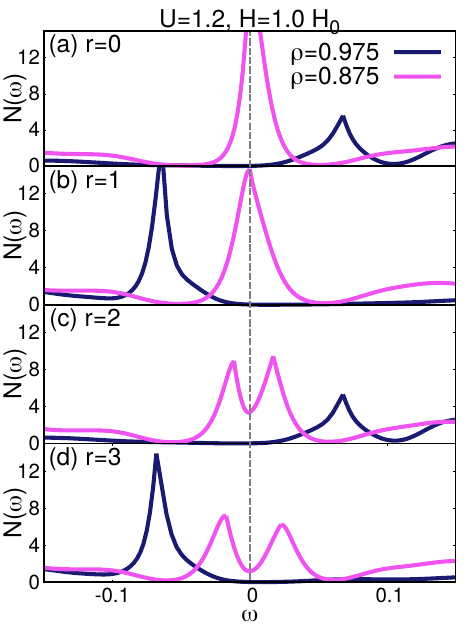}
\caption{
The local density of states $N(r,\omega)$ near the vortex core
for $U=1.2$ (weak coupling) and two electronic densities: $\rho=0.875$ in purple (without CM-cores), and $\rho=0.975$ in blue (for CM-cores). The CdGM-type peaks in blue traces are found at $\omega_0 \approx 0.7\Delta_0$. Here, $r$ is the distance from the center of the vortex. 
The conventional CdGM-peak in LDOS (purple trace) in (a)  splits by moving away from the center of the vortex (a-d). Such splitting is absent for CM-cores, and the location of the corresponding CdGM-type peak (blue trace) oscillates between $\pm \omega_0$ with $r$, signaling CM. 
}
\label{LDOS}
\end{figure}

\vspace{3.5mm}
\noindent \textit{Local density of States (LDOS) at the vortex core} --
Having grasped the spatial modulations in the pairing amplitude and carrier density around vortex cores for a certain parameter space,
We now discuss their consequences on experimentally observable quantities, such as the LDOS $N(r,\omega)$ around a vortex. Here, $r$ represents the distance from the center of the vortex along the $\hat{x}$- or $\hat{y}$-directions. A conventional Abrikosov vortex~\cite{AVL} supports a metallic bound state producing a sharp CdGM-peak at $N(r\approx 0,\omega\approx 0)$~\cite{CdGM},
as shown by the purple trace in Fig.~\ref{LDOS}(a) for $(U,\rho)=(1.2, 0.875)$, for which CM is absent. The corresponding $N(r,\omega)$ for CM-cores is shown by blue traces, representing $(U,\rho)=(1.2, 0.975)$.

While $N(r,\omega)$ in purple traces in Fig.~\ref{LDOS}(a-d) depicts the usual behavior of Abrikosov vortices~\cite{GygiSchluter}, it behaves differently for CM-cores (bllue traces), on three counts: (a) A shift of the CdGM peak to $\omega \approx 0.72\Delta_0$;  (b) No splitting of CdGM peak with increasing $r$; (c) Most importantly, the resulting subgap peak features a ``particle-hole" oscillation, whose location jumps from positive to negative bias with a site-to-site variation of $N(r,\omega)$~\cite{SachdevPRB}. This is due to the underlying charge modulations for which the system prefers inserting or removing electrons from the given site. Such behavior can be picked up in STM experiments~\cite{Hoffman466}. Furthermore, the peak intensity varies weakly with the distance from the core when CM emerges and shows a bias asymmetry with stronger peaks at negative energies than at positive ones.


\begin{figure}[t]
\includegraphics[width=0.49\textwidth]{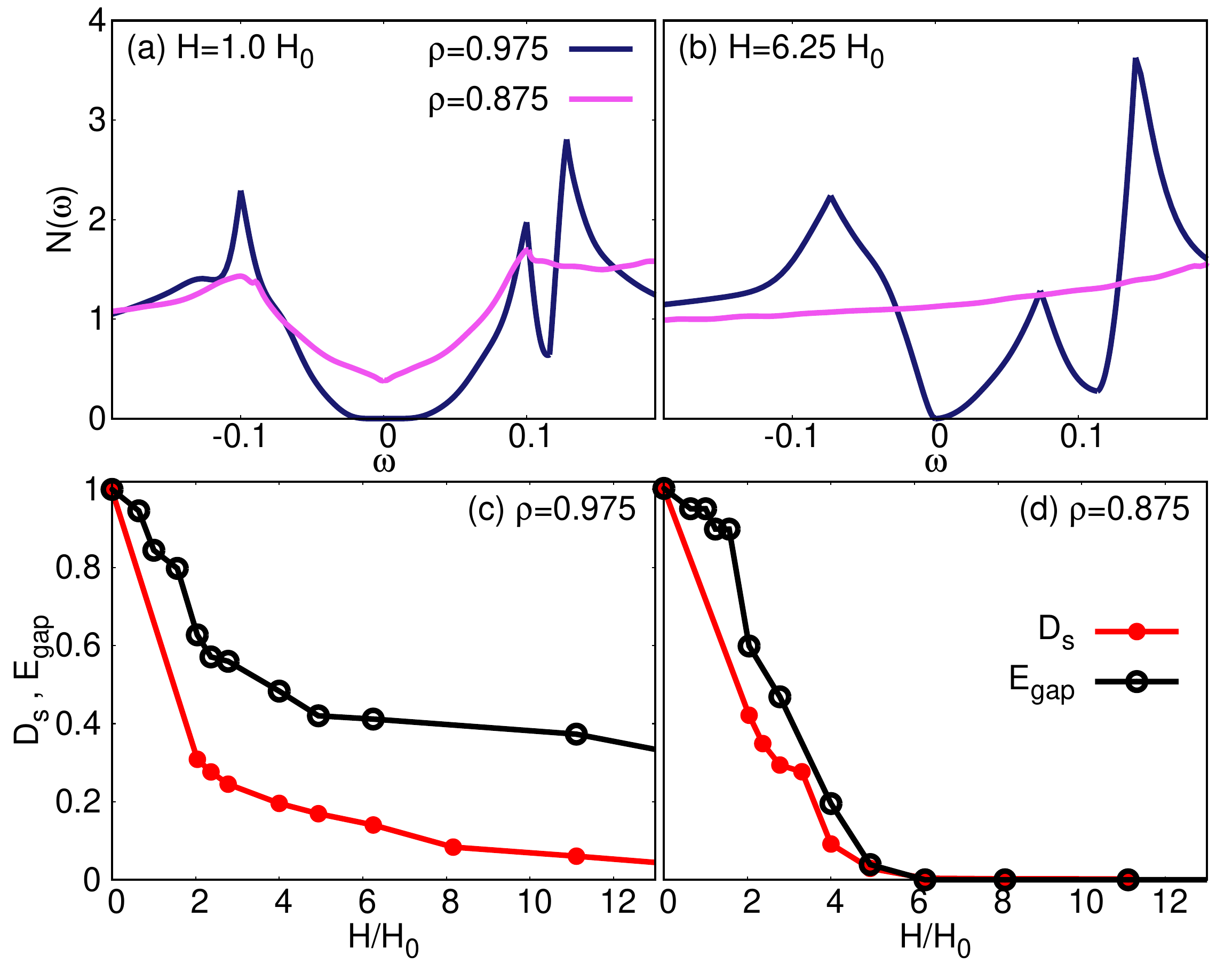}
\caption{
\textbf{(a)} A comparison of the average density of States, $N(\omega)$, for $(U,\rho)=(1.2,0.975)$ by blue trace, and $(U,\rho)=(1.2,0.875)$ by purple trace for $H=H_0$. The $N(\omega)$ at $\rho=0.975$ shows a two-gap structure from SC and CDW ordering. \textbf{(b)} For a much stronger $H=6.25H_0$, the spectral gap in $N(\omega)$ fills up completely for the conventional case ($\rho=0.875$), but it survives for the mixed state with CM-cores ($\rho=0.975$). \textbf{(c)} $H$-dependence of the superfluid density, $D_s$ (red) and the spectral gap $E_{\rm gap}$ (black), normalized by $D_s(H=0)=0.806 \pi$ and $E_{\rm gap}(0)= 0.0895$, respectively, for the case with CM-cores $(U,\rho)=(1.2,0.975)$.\textbf{(d)}, Results similar to panel (b), but for the situation when CM is absent, $(U,\rho)=(1.2,0.875)$. In this case, $D_s(0)=0.802 \pi$ and $E_{\rm gap}(0)=0.0706$.
}
\label{DOS_Ds}
\end{figure}

\vspace{3.5mm}
\textit{Survival of mixed state with CM-cores at large $H$:}
What effects do CM-cores have on the suppression of superconductivity with increasing $H$? We address this by studying the spectral gap in the {\it average} DOS, $N(\omega)$, and superfluid stiffness, $D_s$, as $H$ increases.

Fig.~\ref{DOS_Ds}(a) shows $N(\omega)$ for $(U,\rho)=(1.2,0.875)$ (here, vortex cores are free of CM), as a purple trace. A soft gap in $N(\omega)$ for low fields ($H = H_0$) indicates partial gap filling. In contrast, the blue trace for $(U,\rho)=(1.2,0.975)$ (that produces CM-cores), features a two-gap $N(\omega)$ -- the gap at $\omega \approx 0$ arises from SC~\footnote{Inspite of the partial gap filling, the blue trace of Fig.~\ref{DOS_Ds}(a) features a `hard' SC-gap, quite unlike the conventional case of the purple trace. This threshold energy gap, for the case when CM nucleates in vortex cores is easily comprehended from the structure of LDOS near vortex cores seen in Fig.~\ref{LDOS}.}, while the gap at $\omega = \mu \approx 0.11$ characterizes CM~\cite{AnuragPRB}. A similar two-gap structure of $N(\omega)$ is common to d-wave SC competing with a staggered flux ordering~\cite{KishinePRL}, or with antiferromagnetism~\cite{Kalinsky} in the vortex core.

With an increase in the density of vortices at large $H$, the low-lying core states arising from nearby vortices hybridize for $(U,\rho)=(1.2,0.875)$, promoting dissipative transport which fills up the SC-gap, as shown in Fig.~\ref{DOS_Ds}(b). However, CM-cores for $(U,\rho)=(1.2,0.975)$ strengthen the CDW-gap for the same $H$. Interestingly, the SC-gap, too, survives! A comparison of Fig.~\ref{DOS_Ds}(a, b) demonstrates that the CM in the vortex cores helps maintain the qualitative characteristics of $N(\omega)$ up to a much larger $H$ than when they are absent.
 

The robustness of the vortex phase for the CM-cores is also reflected in the $H$-dependence of $D_s$~\cite{TinkhamBook}, depicted in Fig.~\ref{DOS_Ds}(c) for $U=1.2$. Our findings of $D_s(H)$, normalized by $D_s(H=0)$, for $\rho=0.975$, calculated using the Kubo formalism~\cite{SWZ} shows that $D_s$ decreases with $H$. However, the fall is much slower than for $\rho=0.875$ -- a parameter free of CM-cores (Fig.~\ref{DOS_Ds}(d)). Together with $D_s(H)$, we also present the spectral gap in $N(\omega$), $E_{\rm gap}(H)$, normalized by $E_{\rm gap}(H=0)$ in Fig.~\ref{DOS_Ds}(c, d). Our results illustrate that $D_s$ and $E_{\rm gap}(H)$ evolve hand in hand with $H$, and crash to zero together at $H\approx 5H_0$ for $\rho=0.875$. In contrast, they remain finite even at $H\approx 15H_0$ for $\rho=0.975$. 

This unanticipated ruggedness of the vortex-state with CM-cores is understood as follows:
When the core states are metallic, they are prone to delocalization and hybridization with those from nearby vortices when $H$ enhances, reducing the intervortex spacing. This destroys superconductivity by forming delocalized excitations within the superconducting energy gap. In contrast, the insulating nature of the charge-modulated cores keeps them localized within individual cores, preventing dissipative transport. This, in turn, helps preserve SC-correlations in CM-cores more than when charge modulations are absent. This is elaborated further in the SM~\cite{Supple}.

In summary,
a subdominant charge ordering nucleates in the field-induced vortex cores, for a certain parameter space, conforming vortices to the underlying insulating charge-modulated normal state. Together with the results from Refs .~\cite{Datta_swave,Anushree_dVortex}, above suggests that the vortex-state of an SC outside the conventional limit, where the underlying normal state is not a simple metal, would manifest unusual properties.

\vspace{3.5 mm}
\noindent \textit {Conclusion:} -- 
Going forward it would be useful to investigate the effect of quantum and thermal phase fluctuations of the superconducting order parameter on the vortex cores. It would also be useful to explore the possibility of other CM wave vectors, besides ($\pi,\pi$), as in this work, which is a natural candidate at commensurate density. 

Our main result is that vortex cores reveal the subdominant order in scanning tunneling spectroscopy experiments that can enhance the metrics of a superconductor, such as its critical magnetic field. This opens the door for explorations in other strongly correlated superconductors, such as cuprates, flat band moire, and transition metal dichalcogenides. Interestingly, vortex cores in the SC-state also reveal the underlying normal state, thereby providing a window into a quantum state at a lower temperature that may otherwise not be accessible.

\vspace{3.5 mm}
\noindent \textit {Acknowledgment:} A.B., C.P. and A.G acknowledges funding from CEPIFRA (Grant No. 6704-3). The calculations are performed on the CQM and Dirac clusters at IISER Kolkata. NT acknowledges funding from the seed grant NSF GR137516 as part of the Center for Emergent Materials at Ohio State, an NSF MRSEC, under award number DMR-2011876.
\bibliography{ref}
\bibliographystyle{apsrev4-1}

\clearpage

\onecolumngrid
\vspace{\columnsep}
\begin{center}
\textbf{\large Additional material for ``Charge Modulation in the Vortex Halo of a Superconductor Enhances its Critical Magnetic Field''}
\end{center}
\vspace{\columnsep}
\twocolumngrid

\setcounter{equation}{0}
\setcounter{figure}{0}
\setcounter{table}{0}
\setcounter{page}{1}
\setcounter{enumiv}{0}
\makeatletter
\renewcommand{\theequation}{SM\arabic{equation}}
\renewcommand{\thefigure}{SM\arabic{figure}}
\renewcommand{\theHfigure}{SM\arabic{figure}}
\renewcommand{\thetable}{SM\arabic{table}}
\renewcommand{\bibnumfmt}[1]{[SM#1]}
\renewcommand{\citenumfont}[1]{SM#1}

\subsection{Bogoliubov-de Gennes Formalism}
We perform self-consistent Bogoliubov-de Gennes (BdG) calculations to study the effect of magnetic fields and charge modulations. The mean-field decomposition of the AHM in Eq.~(1) takes the form
\begin{align}
\mathcal{H}_{\rm MF} = 
-t \sum_{\langle i,j \rangle, \sigma} e^{i\phi_{ij}} &\left( c^{\dagger}_{i  \sigma} c_{j \sigma} + \text{h.c.} \right) 
- \sum_{i, \sigma} \left( \mu + \frac{U}{2} \rho_i \right) n_{i \sigma} \nonumber \\
&+ \sum_{i} \left( \Delta_{i} c^\dagger_{i\uparrow} c^\dagger_{i\downarrow} + \text{h.c.} \right),
\label{Comp1}
\end{align}
The local charge density and $s$-wave superconducting (SC) pairing amplitude are self-consistently evaluated 
\[
\rho_i = \sum_{\sigma} \langle c^{\dagger}_{i\sigma} c_{i\sigma} \rangle, \quad
\Delta_i = -U \langle c_{i\downarrow} c_{i\uparrow} \rangle.
\]
where the expectation value is taken over eignestates of $\mathcal{H}_{\rm MF}$.

To diagonalize $\mathcal{H}_{\rm MF}$, we employ the BdG transformation:
\begin{equation}
c_{l \sigma} = \sum_n \left[ \gamma_{n\sigma} u_n(l) - \sigma \gamma^\dagger_{n \bar{\sigma}} v_n^*(l) \right],
\label{bdg11}
\end{equation}
which leads to the BdG eigenvalue equations
\[
\begin{pmatrix}
    \hat{\xi} & \hat{\Delta} \\
    \hat{\Delta}^* & -\hat{\xi}
\end{pmatrix}
\begin{pmatrix}
    u_n(l) \\
    v_n(l)
\end{pmatrix}
= \epsilon_n
\begin{pmatrix}
    u_n(l) \\
    v_n(l)
\end{pmatrix}.
\]
with
\begin{align}
\hat{\xi} u_n(l) &= -\sum_{\delta} t_\delta u_n(l+\delta) - \left( \mu + \frac{U}{2} \rho_l \right) u_n(l), \label{bdg16} \\
\hat{\Delta} u_n(l) &= \Delta_l v_n(l), \label{bdg17}
\end{align}
with $t_\delta = t e^{i\phi_{i,i+\delta}}$ and $l=1,\hdots,N_s$ is the total number of sites and the $n=1,\hdots,N_s$ is the  total number of eigentates, and $\hat{\xi}$ and $\hat{\Delta}$ are thus $N_s \times N_s$ matrices.

We model CDW order using the ansatz
$\rho_i = \rho + \Delta_{\rm CDW}(i) e^{i \mathbf{Q} \cdot \mathbf{r}_i}$, with $\mathbf{Q} = (\pi, \pi)$
where the local modulation amplitude is given by $\Delta_{\rm CDW}(i) = e^{-i \mathbf{Q} \cdot \mathbf{r}_i} (\rho_i - \rho)$
and the corresponding CDW energy scale is defined as $\tilde{\Delta}_{\rm CDW} =(U/2))\Delta_{\rm CDW}$.

We exploit translational symmetry of the vortex lattice by solving $\mathcal{H}_{\rm MF}$ on a magnetic unit cell, extending the solution over typically $10 \times 10$ supercells using repeated zone schemes~\cite{Kalinsky_SM}. The total number of flux quanta is chosen to be even to accomodate periodic boundary conditions. To accelerate convergence of the self-consistency loop, we employed linear, the modified Broyden mixing~\cite{ModBroyd_SM}.

We focus on two key physical observables:
\begin{itemize}
    \item[1.] The Local density of states (LDOS) at site $r$ and energy $\omega$ is given in terms of BdG eigenvectors as
\begin{equation}
N(r, \omega) = \sum_n \left[ |u_n(r)|^2 \delta(\omega - \epsilon_n) + |v_n(r)|^2 \delta(\omega + \epsilon_n) \right]
\end{equation}
whereas the average density of states is obtained by spatial averaging $N(\omega) = \frac{1}{N_s} \sum_r N(r, \omega)$.
\item[2.] Superfluid stiffness, calculated within linear response theory, is expressed via the Kubo formula as~\cite{SWZ_SM}
\begin{equation}
\frac{D_s}{\pi} = \langle -K_x \rangle + \Lambda_{xx}(q_x = 0, q_y \rightarrow 0, \omega = 0),
\end{equation}
where $K_x$ is the kinetic energy along the $x$-direction, and $\Lambda_{xx}$ denotes the paramagnetic current-current correlation function in the static, long-wavelength limit~\cite{SWZ_SM}.
\end{itemize}

\subsection{Determining length scales $\xi_{\rm SC}$ and $\xi_{\rm CM}$}
This section describes extraction of the  charge modulation (CM) and the superconducting (SC) length scale from the local electronic density and  the pairing amplitude respectively. 

\begin{figure}[t]
\includegraphics[width=0.48\textwidth]{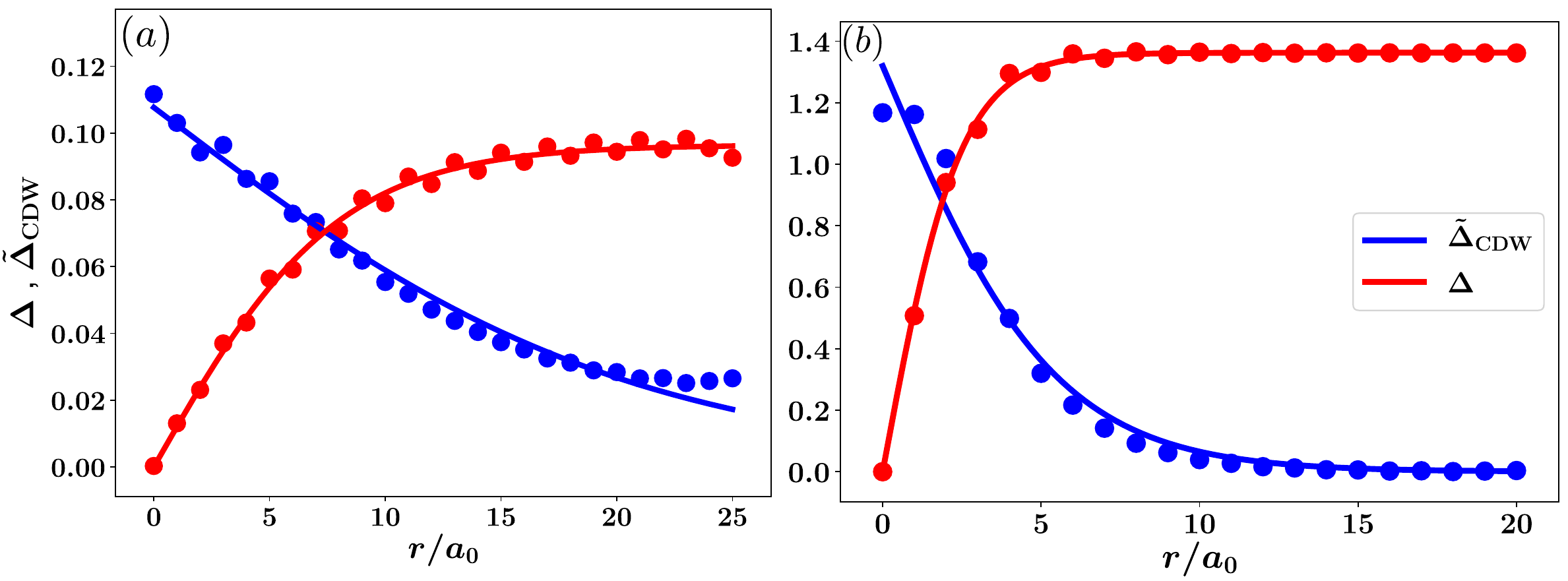}
\caption{The superconducting $\Delta$ and  $\tilde{\Delta}_{\rm CDW}$ as a function of $r/a_0$ averaged over the azimuthal direction. (a) For $(U,\rho)=(1.2,0.975)$ (b) For $(U,\rho)=(4.0,0.875)$. The extracted length scales are tabulated in Table~\ref{Table1}.}
\label{fig:figApp1}
\end{figure}

The profiles ($\Delta$ and $\tilde{\Delta}_{\rm CDW}$) are averaged over the azimuthal direction and plotted as a function of radial distance from the vortex core in Fig.~\ref{fig:figApp1}. As expected, the SC vanishes inside the core regions, where CM amplitude is strongest. Such dependence suggests a  $\tanh$ functional used in the past to to extract  SC coherence length $\xi_{\rm SC}$~\cite{GygiSchluter_SM}. Since the CDW order shows qualitatively complementary spatial behavior compared to SC pairing, we use a corresponding $\tanh$ function to fit the two radial profile,
\begin{align}
    \Delta(r)&=\Delta^{(0)} \tanh\left(r/\xi_{\rm SC}\right) ,\\
    \tilde{\Delta}_{\rm CDW}(r)&= \Delta^{(0)}_{\rm CDW} \left[1-\tanh\left(r/\xi_{\rm CM}\right) \right].
\end{align}
We have quoted the values of $\xi_{\rm SC}$ and $\xi_{\rm CM}$ in Table~\ref{Table1} for different values of $(U,\rho)$. We have also listed the uniform values of $\Delta_0$ and the subdominant $\tilde{\Delta}_{\rm CDW}$. The coefficients $\Delta^{(0)}$ and $\Delta^{(0)}_{\rm CDW}$ closely match $\Delta_0$ and $\tilde{\Delta}_{\rm CDW}$ presented in Table~\ref{Table1}. While the exact values of the length scales depend on the functional form, the relative behavior of $\xi_{\rm SC}$ and $\xi_{\rm CM}$ remains broadly consistent across different functions.

According to Landau Ginsberg formalism, the coherence length increases for weaker couplings and  $\xi \sim 1/\phi_0$, where $\phi_0$ is the order parameter strength of the uniform system. Such a feature is roughly maintained in our estimations for both orders. Since CDW is the subdominant order, it shows a larger length scale than the SC. However, the presence of the CDW also weakens the pairing near the vortex core regions, and hence, the SC coherence length scale with CM-cores for $(U,\rho)=(1.2,0.975)$ is almost double than that in the metallic cores for $(U,\rho)=(1.2,0.875)$. 

\begin{table}[]
\begin{tabular}{|c|c|c|c|c|}
\hline
$(U,\rho)$    & $\xi_{\rm SC}/a_0$  & $\Delta_0$ & $\xi_{\rm CM}/a_0$     & $\tilde{\Delta}_{\rm CDW}$ \\ \hline
$(1.2,0.875)$ & $4.13 \pm 0.156$ & $0.071$    & ---                & ---           \\ \hline
$(1.2,0.975)$ & $7.85 \pm 0.094$ & $0.101$    & $22.50 \pm 2.017$ & $0.044$       \\ \hline
$(3.0,0.875)$ & $2.86 \pm 0.445$ & $0.830$    & $5.47 \pm 0.090$  & $0.417$       \\ \hline
$(4.0,0.875)$ & $2.83 \pm 0.372$ & $1.362$    & $5.51 \pm 0.067$  & $0.912$       \\ \hline
\end{tabular}
\caption{Extracted charge modulation length $\xi_\chi$ and superconducting length scale $\xi_\Delta$ for different parameters. Along with the uniform value of superconducting pairing $\Delta_0$ and charge modulation energy $\tilde{\Delta}_{\rm CDW}=(U/2) \Delta_{\rm CDW}$ where $\Delta_{\rm CDW}$ is the uniform modulation amplitude of the electronic density.} 
\label{Table1}
\end{table}

\subsection{Inhomogeneous charge modulations in the normal state}
\begin{figure}[t]
\includegraphics[width=0.48\textwidth]{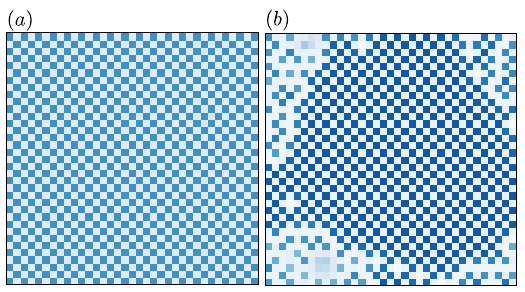}
\caption{Depicts the spatial profile of self-consistent local densities for the normal state at $(U,\rho)=(4.0,0.875)$. \textbf{(a)} Displays the uniform CDW state as one of the candidates of the normal state, with energy per site $F=-2.415t$ \textbf{(b)} Shows a representative inhomogeneous charge density modulating phase with $F=-2.487t$, thus energitically favorable than the uniform CDW state. In these profiles, puddles of regions with density modulations around $\rho=1$ develop spontaneously. Some spatial regions go below $\rho$ to maintain the average density, creating a heterogeneous density profile.}
\label{fig:figApp2}
\end{figure}
In the main text, we mentioned that the CM-cores lead to an accumulation of electron density close to half-filling at the vortex regions. For example, our estimate of the average density in a $4\times 4$ patch, $\varrho^{(4)}(U,\rho)$, surrounding a vortex for ($U=4.0,\rho=0.875$), $\rho^{(4)}(4.0,0.875) \approx 0.944$ and for ($U=4.0,\rho=0.875$), is $\rho^{(4)}(4.0,0.875) \approx 0.990$. However, we have $\rho^{(4)}(1.2,0.875) \approx 0.875$ for the uniform vortex core. Thus, the electronic density modulates in the core region around half-filling. 

The s-wave superconductivity and CDW with wavevector, $\bf{Q}=(\pi,\pi)$, constitute a degenerate ground state of AHM at half-filling for clean systems~\cite{MicnasRMP_SM}. We address here whether such modulation survives away from half-filling, if so,  we address the spatial features of electron density for $\rho<1$.

To investigate the stability of the uniform CDW state (shown in Fig.~\ref{fig:figApp1} (a)) against inhomogeneous charge modulations (ICM), we performed a self-consistent estimation of the normal state starting with different independent initial density profiles. We observe a striking behavior in several quasi-degenerate ICM states that lowered the energies compared to the uniform CDW. One such spatial profile of representative densities is shown in Fig.~\ref{fig:figApp1} (b) for $(4.0,0.875)$. A common feature of all such ICM states also anchors the average density close to half-filling. This reduces the density of the ``domains" between ICM patches. Such domains maintain the set average electronic density. Our estimation of the phase boundary in main text Fig.~2 (a) considers Uniform CDW as a trial and error scheme to determine the phase boundary of ICM ground state, is inadequate and computationally costly. Our findings hint that such ICM states encroach part of the uniform phase diagram upon considering ICM solutions.

Our exploration of ICM patterns in the normal state away from half-filling reveals a simple picture. Periodic charge modulations arise due to:  
(a) On-site attraction, which promotes double occupancies, and  
(b) Virtual electron hopping enhances kinetic energy $E_t \propto t^2/U$ when sites adjacent to double occupancy remain empty~\cite{RTS_SM}, favoring kinetic energy gain. This mechanism is most effective at $\rho = 1$. Perfect charge modulations cannot persist when deviating from half-filling, making superconductivity the preferred ground state. However, when superconductivity is suppressed, finite regions at half-filling can still sustain perfect charge ordering, separated by domains with $\rho \leq 1$.  The overall kinetic energy minimizes when ICMs aggregate into fewer but larger clusters at a given $\rho$, reducing the energy cost associated with `surface tension' at domain walls. As doping increases, non-ICM regions expand, eroding the energy advantage and favoring a uniform-density ground state.

\begin{figure}[t]
\includegraphics[width=0.48\textwidth]{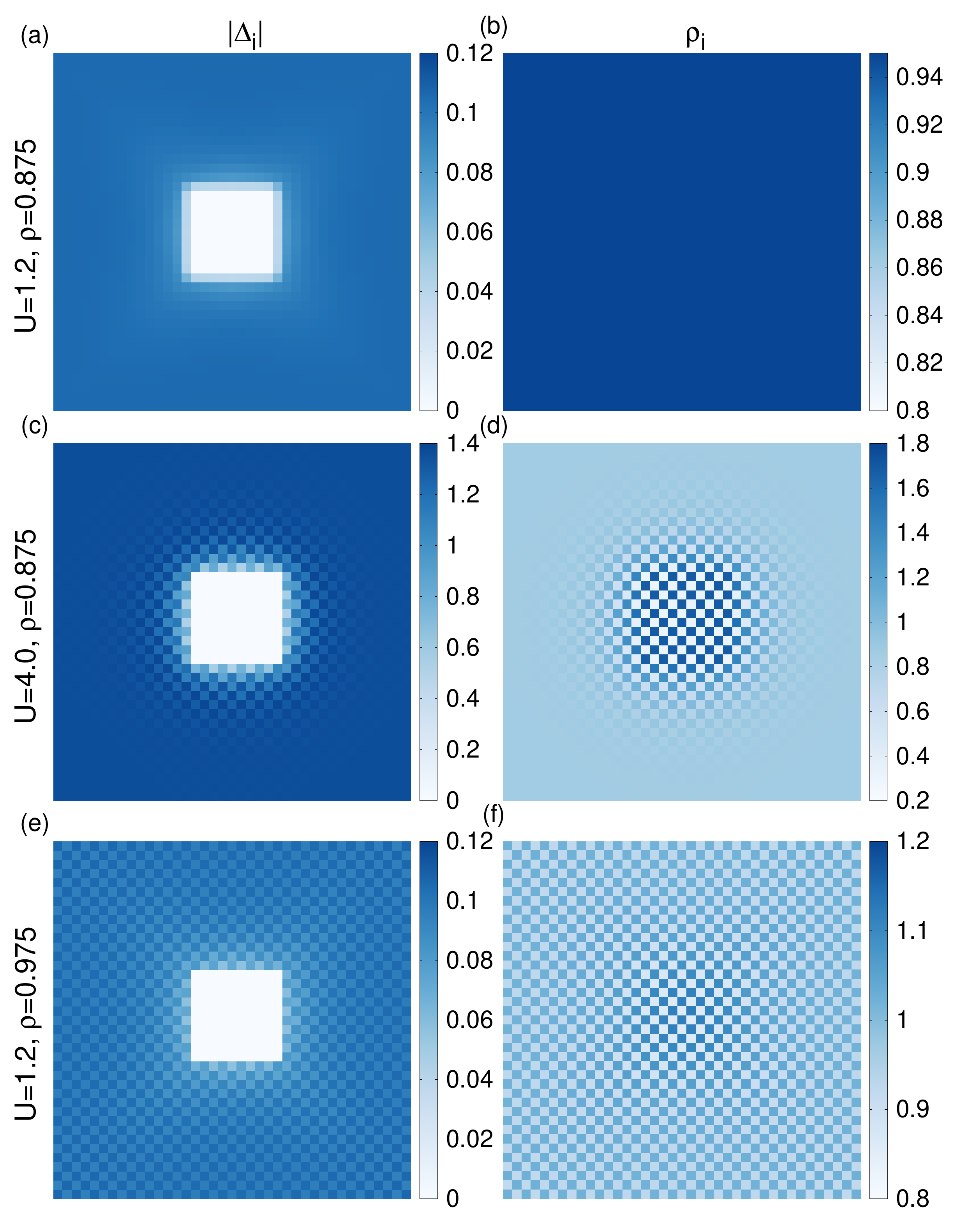}
\caption{The left panels depict the self-consistent SC pairing amplitude in a system where pairing is not allowed in the central region, creating a local normal region. \textbf{(a)} For $(U,\rho)=(1.2,0.875)$ \textbf{(c)} For $(U,\rho)=(4.0,0.875)$ \textbf{(f)} For $(U,\rho)=(1.2,0.975)$. The right panels present self-consistent local density for the corresponding parameters on the left. The local suppression of $\Delta_i$ in the central regions helps nucleate charge modulations in those regions if the underlying normal state allows it. The central regions become half-filled whenever charge modulations emerge there.}
\label{fig:figApp3}
\end{figure}
\subsection{Emergence of half-filled regions and its connection to CM}
In this section, we show that the CM emerges in the regions where SC is suppressed by creating a half-filled region by performing an independent calculation. We set the $\Delta_i=0$ in a central $10\times10$ region in a $40\times 40$ system. For these sites, $\Delta_i$ is not evaluated self-consistently and is fixed at zero, whereas standard self-consistency protocol for $\Delta_i$ is followed in the rest of the system. The local density is evaluated self-consistently everywhere in the system. This forced suppression of SC helps us identify if the charge modulations discerned in the normal state can emerge locally for the favorable parameter range in the absence of magnetic field. Our findings are presented in Fig.~\ref{fig:figApp3}.
 
This analysis shows that regions where local SC pairing is eradicated support charge modulation if the underlying normal state allows it. The charge modulations fix the average density to half-filling in regions of suppressed superconductivity. The magnetic field acts as a perturbation to suppress superconductivity in the vortex area, enabling the normal state to manifest naturally. However, the vortex core eradicates superconducting pairing only at a single site, as the sSC pairing recovers over the coherence length $\xi_{\rm SC}$. Consequently, the average density in the core regions is influenced by the presence of finite pairing in nearby areas. In the main text, the CM core regions also tend to form such half-filled patches.

\subsection{Participation ratio of the low-energy excitations}
\begin{figure}[t]
\includegraphics[width=0.48\textwidth]{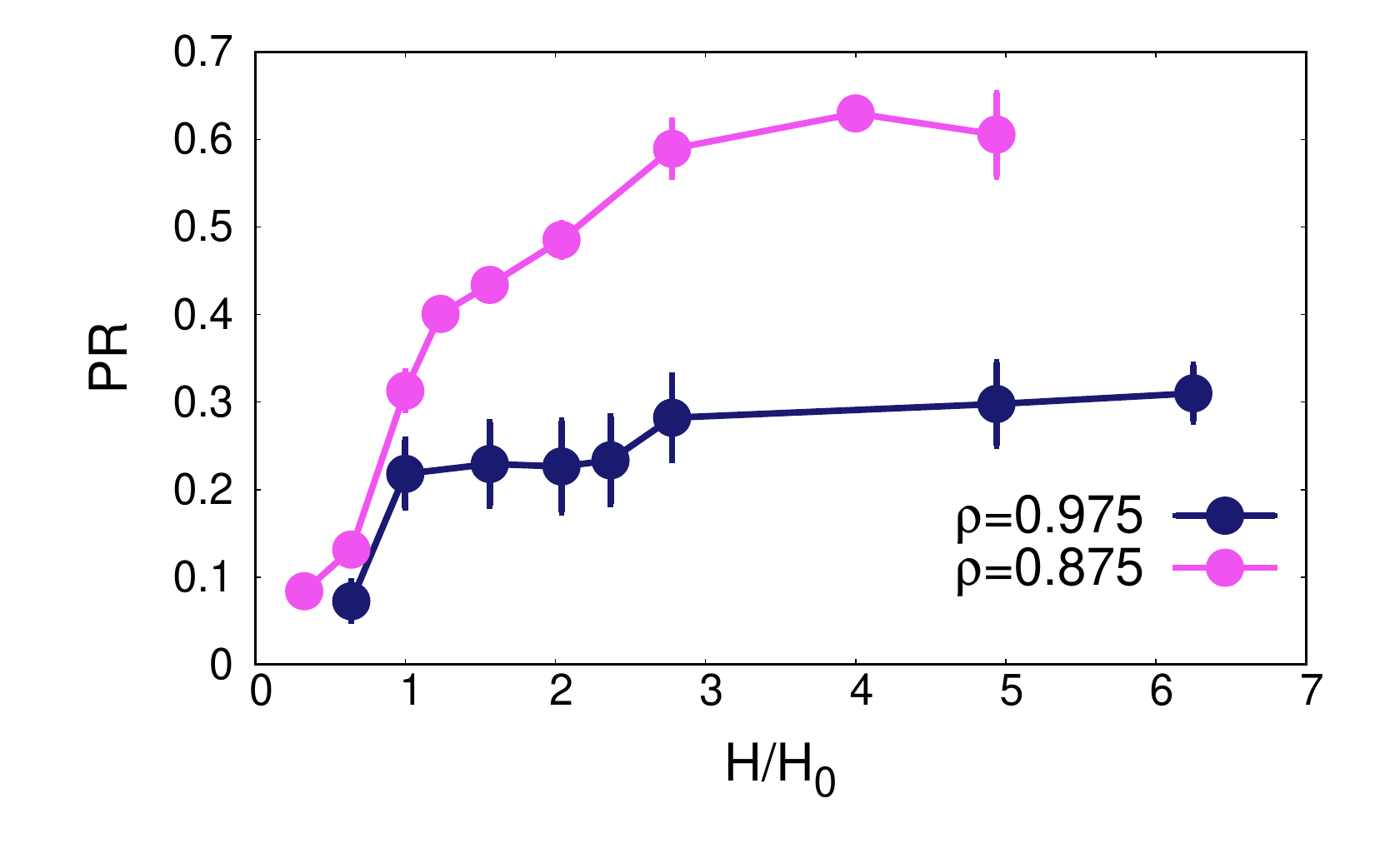}
\caption{Exhibits the evolution of the participation ratio of the five
lowest excitations for CM-cores $(U,\rho)=(1.2,0.975)$ and metallic cores $(U,\rho)=(1.2,0.875)$. The CM core's insulating nature helps keep the low-energy states bound to it. Increasing fields brings these bound states
closer, leading to hybridization, which kills superconducting correlations. However,  a lower PR for a modulated core shows a weaker hybridization propensity, protecting superconductivity up to the larger magnetic field.}
\label{fig:figApp4}
\end{figure}

In this section, we discuss the rationale for the stability of the superfluid stiffness with a magnetic field for charge modulated vortex core. For a conventional vortex core (in the absence of other fluctuations), the destruction of superconductivity with increasing magnetic field can be understood by the overlapping bound states from nearby vortices, leading to quasi-bound and eventually extended states. This mechanism is analogous to the hybridization of localized atomic orbitals in tight-binding models~\cite{CanelPL_SM}. 

To capture such localization-to-delocalization behavior, we study the participation ratio (PR) of the sub-gap states. The PR is directly proportional to the spread of the wavefunction over the system of the low-energy states and is given by
\begin{equation}
\zeta_{n}=\frac{1}{N}\sum_{i=1}^{N} \frac{1}{\vert\Psi_{n}(\mathbf{r}_i)\vert^4},
\label{eqn:PR}
\end{equation}
where, $\zeta_{n}$ is the participation ratio of the nth eigenstate. PR tend to one if the eigenstate is completely delocalized and should approach $\sim 1/N$ if the wavefunction localizes on a single site. We have averaged over the PR for the lowest five excitations.

We compare the $H-$evolution of participation ratio in Fig.~\ref{fig:figApp4} for $U=1.2$ for CM and metallic vortex cores. The magnetic field increases the participation ratio, indicating the delocalization of the core states in both cases. However, the core states remain relatively localized when charge modulations nucleate in the core regions. Such localized states prevent the hybridization of the nearby core states, anchoring the core states to a particular vortex. This prevents the destruction of the superfluid density of the superconductor, as seen in the main text.

\end{document}